\documentclass[aps, twocolumn]{revtex4}

\usepackage{float}
\usepackage{dcolumn}
\usepackage{amsmath}
\input{epsf}
\input{epsfx}

\ifx\pdftexversion\undefined
  \usepackage{graphicx}
\else
  \usepackage[pdftex]{graphicx}
\fi
\usepackage{epsfig}
\usepackage{ulem}

\begin{document}
\title{Local temperature control of photonic crystal devices via micron-scale electrical heaters}

\author{Andrei Faraon \footnote[1] {Electronic address: faraon@stanford.edu }, Jelena Vu\v{c}kovi\'{c}}
\affiliation{E. L. Ginzton Laboratory, Stanford University, Stanford, CA, 94305, USA}


\begin{abstract}

We demonstrate a method to locally control the temperature of photonic crystal devices via micron-scale electrical heaters. The method is used to control the resonant frequency of InAs quantum dots strongly coupled to GaAs photonic crystal resonators. This technique enables independent control of large ensembles of photonic devices located on the same chip at tuning speed as high as hundreds of kHz.

\end{abstract}

\maketitle

Integrated devices composed of optical resonators and waveguides are seen as one of the most promising solutions for future optical networks for information processing\cite{06NodaPCreview}. One of the main drawbacks of using resonators is that their frequencies are highly sensitive to fabrication errors and any fluctuations in the environment that leads to changes in the index of refraction or device geometry. The complexity of designing these systems increases even more when resonators need to be coupled to single optical emitters, as is the case for devices used in quantum information science. One possible solution is to develop system capabilities where the resonators can be controlled reversibly and on an individual basis. Some of the most scalable methods for local tuning is by controlling the index of refraction via carrier injection\cite{2008.Finley.ElectricalControlSC,2005.LipsonSiliconElectroOpticsModulator} or by changing the temperature\cite{AndreiTtune,2006.Panepucci.MicroRingTTune}. Generally, the local control of temperature is slower than the carrier recombination, so it is preferable to use temperature for device tuning and carrier control for other functions, like optical switching. We have already reported a method to locally control the temperature of photonic crystal devices by heating using a laser beam\cite{AndreiTtune}. This method allows for reliable control of individual devices located on the same chip, but for large networks of resonators it is preferable for the tuning to be done electrically. In this letter we demonstrate local tuning of photonic crystal devices using micron-scale electrical heaters. Although this technique primarily targets the tuning of quantum dots(QDs) and photonic crystal cavities in optical networks for quantum information science, the method can be applied to any type of optical networks where local control of temperature is desirable for tuning purposes\cite{2007.ReconfigurableMicroringsThermallyTuned}.

\begin{figure}[htbp]
    \includegraphics[width=3.5in]{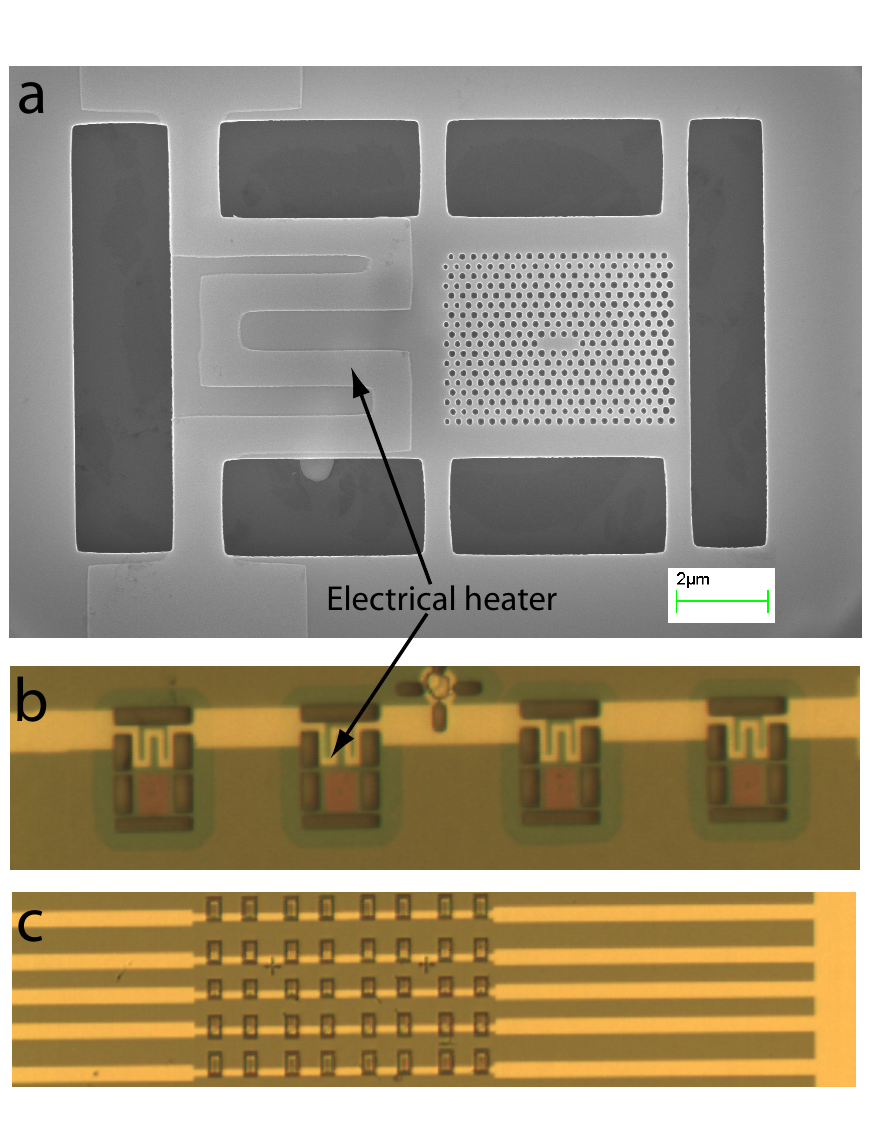}
\caption{(a) Scanning electron microscope image of the suspended GaAs photonic crystal cavity and the heating pad. The serpentine pattern on the heating pad is the Ohmic heater. (b,c) Optical microscope images showing multiple devices interconnected electrically.}
\label{fig:device}
\end{figure}

To integrate local heaters with the QD/photonic crystal structures, we designed the device shown in Fig.\ref{fig:device}(a). The concept of the device is similar to the one in Ref.\cite{AndreiTtune}, and it consists of a GaAs photonic crystal cavity next to a heating pad, both suspended via six narrow bridges to increase the thermal insulation. The device is fabricated in a 160nm thick GaAs membrane with a mid-layer of InAs quantum dots, grown on top of a $1 \mu m$ thick AlGaAs slab that can be wet etched in hydrofluoric acid. The photonic crystal and the heating pad cover a $11 \mu m \times 5 \mu m$ area. The left suspension bridges in Fig.\ref{fig:device}(a) were $2 \mu m$ long, $1 \mu m$ wide, while the other were $2 \mu m$ long, $0.55 \mu m$ wide. The chip was fabricated using electron-beam lithography, dry plasma etching, metal deposition and wet etching of the sacrificial layer.

The temperature control is achieved via micron-scale ohmic heaters located on the heating pad. The heaters were first patterned by electron-beam lithography and then two metal layers (15nm Au on top of 20 nm Cr) were deposited. A serpentine shape was chosen in order to increase the resistance of the device. Multiple devices on the same chip can be electrically connected\ref{fig:device}(b,c). The devices were connected in series in groups of eight, four devices being shown in Fig.\ref{fig:device}(b). Twenty sets were connected in parallel to two bonding pads that could be connected to an electrical power supply. The total resistance of the chip was $\sim 80 \Omega$, implying a resistance of $\sim 200 \Omega $ for each individual heater.

\begin{figure}[htbp]
    \includegraphics[width=3.5in]{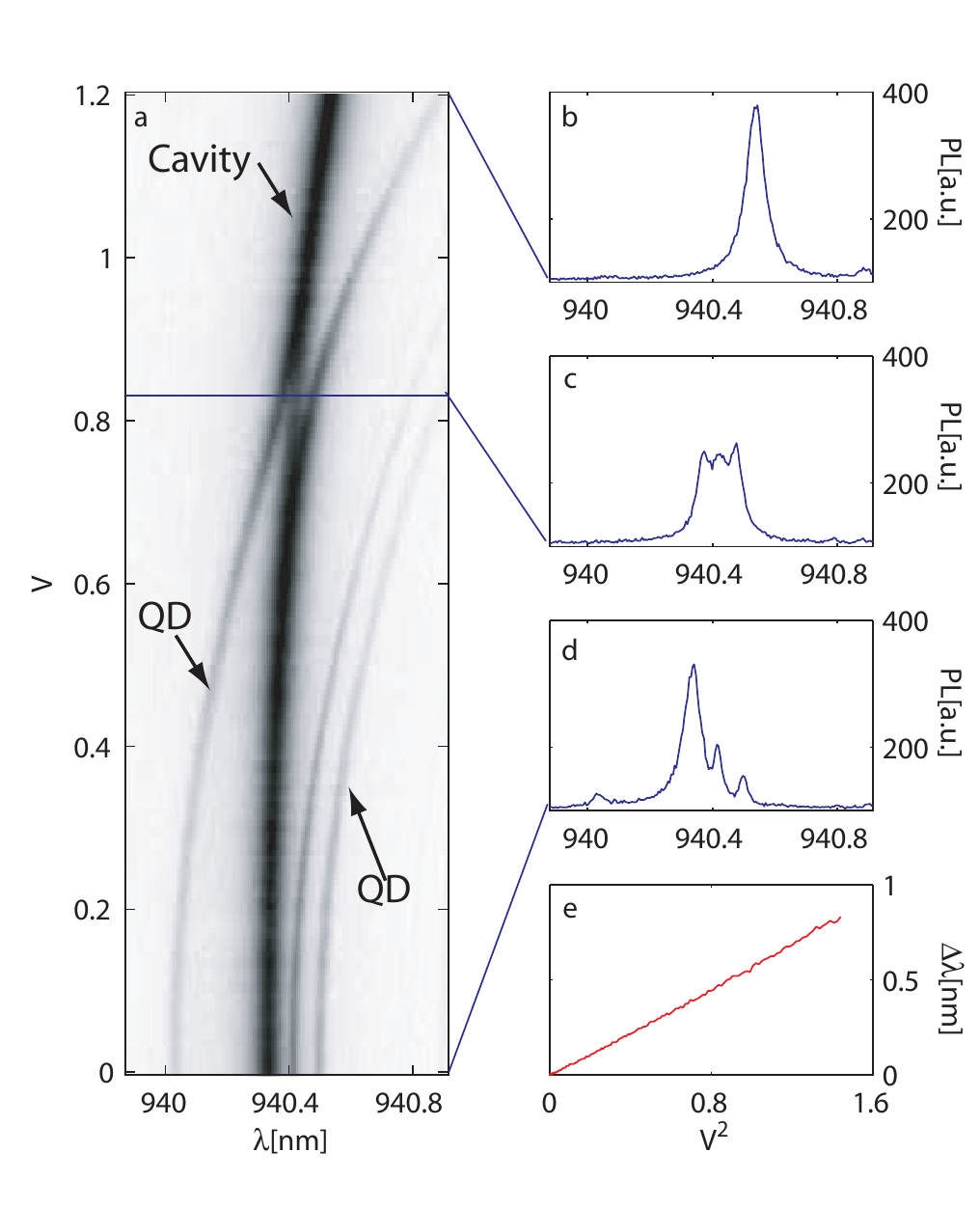}
\caption{(a-d) Quantum dot and cavity tuning performed by controlling the temperature of the device using local electrical heating. Anti-crossing, is observed as the quantum dot is tuned into resonance with the cavity, a signature of the strong coupling regime. (e) Linear dependence of the shift in the quantum dot frequency with the square of the applied voltage.}
\label{fig:SC_electrical}
\end{figure}

The device was cooled down in a He flow cryostat and kept at a base temperature of 5K, such that photoluminescence from single quantum dots could be observed. The measurements were performed using a confocal microscope setup and a laser tuned above the band gap of GaAs to excite photoluminescence\cite{NatureRef}. Electrical control was achieved by connecting the entire chip to a function generator. The photoluminescence from a coupled cavity/QD system was monitored while the voltage was linearly changed from 0V to 1.2V, as shown in Fig.\ref{fig:SC_electrical}(a-d). Red shift of the QD and cavity frequency is observed while increasing the bias voltage. Both the cavity and the quantum dot frequency shift show a quadratic dependence on the applied bias voltage. Fig.\ref{fig:SC_electrical}(e) shows the linear shift in the quantum dot with the square of the driving voltage(proportional to the dissipated power),  in agreement with previous results in Ref.\cite{AndreiTtune}. To obtain the tuning range shown in Fig.\ref{fig:SC_electrical}, corresponding to a temperature increase from $5K$ to $\sim 25K$, a thermal power of $\sim 0.1 mW$ was dissipated in the device.

Anti-crossing was observed as one of the quantum dot lines is tuned into resonance with the cavity, thus indicating strong coupling\cite{Yoshie04,AndreiTtune}. A fit to the data indicated a cavity quality factor $Q \sim 13000$ and a QD/cavity coupling rate $g/2\pi \sim 17GHz$. For this particular cavity/QD system, a triplet of spectral lines was observed when the cavity is tuned into resonance with the quantum dot. The side spectral lines correspond to the cavity/QD polaritons, while the middle spectral line should correspond to the cavity being populated with photons from other neighboring quantum dots\cite{SCImamogluNature}.

\begin{figure}[htbp]
    \includegraphics[width=3.5in]{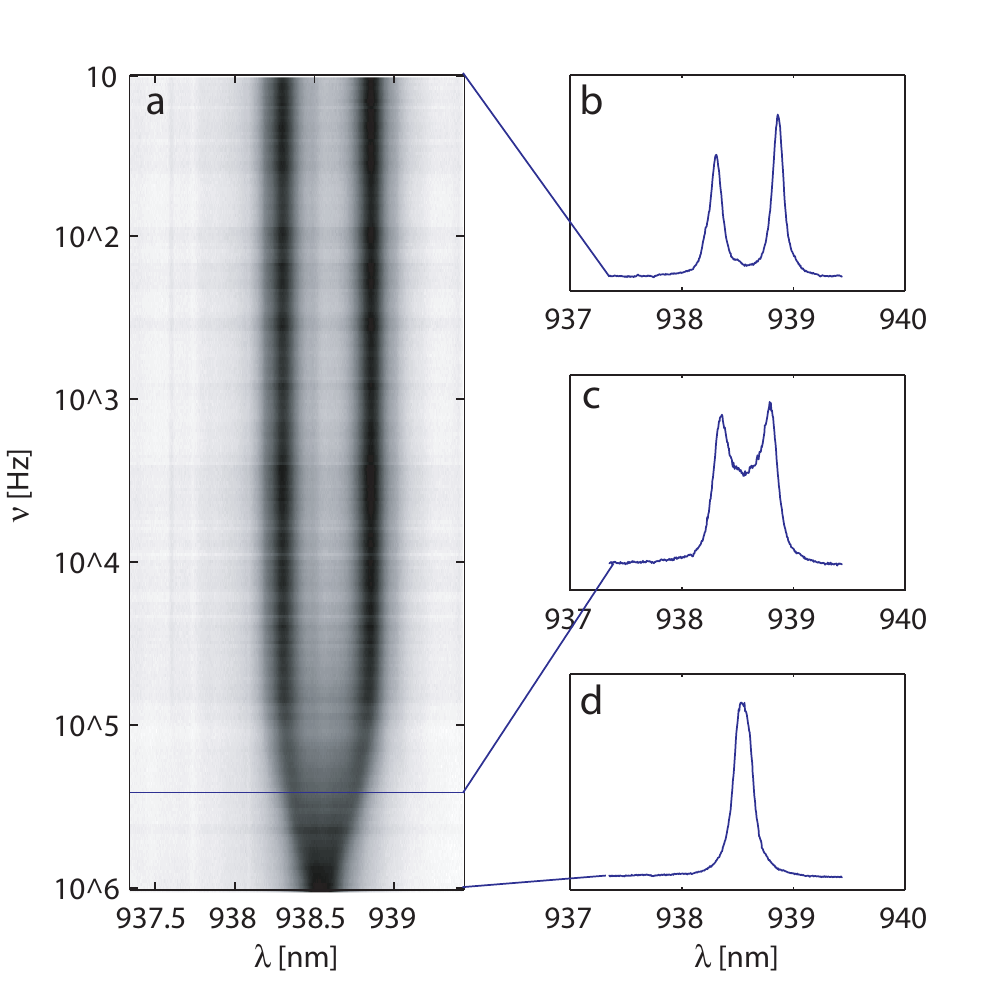}
\caption{(a)Photoluminescence spectra integrated over one second as the temperature is tuned via a square wave signal from 10Hz to 1MHz. (b-d) Spectra showing three different regimes of operation. (b)Thermal relaxation rate ($\Gamma$) much faster than the driving frequency($\omega$). (c)$\Gamma \sim \omega$. (d) $\Gamma < \omega$. }
\label{fig:speed}
\end{figure}

For some applications, where fast fluctuations of the environment cause rapid changes in the cavity resonance, the maximum tuning speed is a relevant parameter. The maximum speed achievable with this tuning technique is limited by the thermal relaxation time of the device $\tau$. The thermal relaxation time was inferred by measuring the frequency response. A device similar to the one shown in Fig. \ref{fig:device}(a) but without coupled quantum dots was used for this measurement. The system was driven using a square wave (0V-2V) and the frequency was continuously swept from 10Hz to 1MHz. At the same time, the photoluminescence spectrum was monitored on a spectrometer at a refresh rate much slower than the modulation frequency. For modulation rates much slower than the thermal relaxation rate, the temperature of the device could follow precisely the square wave form of the driving signal, so the temperature was either $T_{0} \sim 5K$ or $T_{1} \sim 60K$ corresponding to 0V or 2V. The spectrometer showed two resonances, corresponding to the cavity frequency for the two different temperatures ($T_{0}$ and $T_{1}$). As the frequency is increased beyond the thermal relaxation rate, the temperature of the device could not follow the electrical modulation and it stabilized at the temperature corresponding to the average dissipated power. The experimental data is shown in Fig.\ref{fig:speed} indicating that the device can be driven up to $\sim 100 KHz$ ($\tau \sim 10\mu s$). Depending on the application the speed of the device can be increased by using thicker suspension bridges thus increasing thermal conductivity. However, this comes at extra energy cost since more power needs to be dissipated in order to keep the device temperature at the desired level. For some applications the use of suspended structures may not be required. This could be the case for silicon on insulator devices, where the silicon oxide substrate has low thermal conductivity.

In conclusion, we show that the temperature of suspended photonic crystal devices can be controlled electrically via micron-scale electrical heaters. This would enable efficient tuning of large ensembles of photonic crystal devices located on the same chip. Tuning speeds up to 100KHz are demonstrated, limited by the thermal relaxation time of the device.


Financial support was provided by the ONR Young Investigator Award, Presidential Early Career Award, Army Research Office. Part of the work was performed at the Stanford Nanofabrication Facility of NNIN supported by the National Science Foundation. The authors thank Prof. Pierre Petroff and Dr. Nick Stoltz for providing the GaAs material.


\bibliographystyle{unsrt}

\end{document}